# Predicting Engagement with the Internet Research Agency's Facebook and Instagram Campaigns around the 2016 U.S. Presidential Election

Dimitra (Mimie) Liotsiou[1*], Bharath Ganesh[1,2], Philip N. Howard[1*]

**Affiliations:**

[1] Oxford Internet Institute, University of Oxford, Oxford, UK.

[2] Faculty of Arts, University of Groningen, Groningen, the Netherlands.

* Corresponding authors. Email: {mimie.liotsiou, philip.howard}@oii.ox.ac.uk

**Abstract**

The Russian Internet Research Agency's (IRA) online interference campaign in the 2016 U.S. presidential election represents a turning point in the trajectory of democratic elections in the digital age. What can we learn about how the IRA engages U.S. audiences, ahead of the 2020 U.S. presidential election? We provide the first in-depth analysis of the relationships between IRA content characteristics and user engagement on Facebook and Instagram around the 2016 election. We find that content targeting right-wing and non-Black marginalised groups had the strongest positive association with engagement on both Facebook and Instagram, in contrast to findings from the IRA campaign on Twitter and to some previous commentary in the media. Higher engagement was associated with posting later in the 2015-2017 period and using less text on both platforms, using negative wording and not including links on Facebook, and using fewer hashtags on Instagram. The sub-audiences and sub-issues associated with most engagement differed across the platforms.

**Introduction**

In recent years, the proliferation of disinformation and propaganda campaigns on social media around elections and important political events has been recognized as an important threat to democracy internationally. In this context, the Russian Internet Research Agency's (IRA) interference campaign in the 2016 U.S. presidential election has been viewed as a key turning point. There is significant evidence that Russia's direct engagement with U.S. voters has intensified rather than abated since then [1, 2, 3]. For social media platforms, preventing foreign election interference continues today to pose a crucial challenge [4], which has been characterised as an 'arms race' by the CEO of Facebook (which owns the Facebook and Instagram platforms) [4, 5]. In anticipation of the next U.S. presidential election in 2020, we focus on what can be learnt about how Russia's IRA engages with U.S. audiences, using unique fully attributed data from the IRA's social media interference campaign around the 2016 U.S. election.

In 2017, the U.S. Senate Select Committee on Intelligence (SSCI) compelled three big tech companies (Facebook, which owns the Facebook and Instagram platforms; Twitter; and Google, which owns the YouTube platform) to provide to them data on the IRA's interference operations on their platforms. Some of this data has been made publicly available (Twitter posts, Facebook ads), but a larger dataset including also non-public data was given to external independent researchers, resulting in two independent reports published in December 2018 [6, 7]. These reports provided initial analyses of the characteristics, spread and dynamics of the IRA campaign, as well as the communities and issues targeted by it, on all of the above platforms.

The above reports did not investigate whether user engagement with IRA content could be predicted based on the features of the IRA posts, nor whether and what kinds of relationships exist between IRA content features and user engagement. We offer the first in-depth investigation of these questions, using non-public data from the SSCI dataset on the IRA campaigns run on Facebook and Instagram over 2015-2017.

Media commentary on the above two reports [6, 7] often focused particularly on the IRA's targeting of Black communities, including coverage by major U.S. and U.K. news outlets [8, 9, 10, 11, 12, 13]. Additionally, a study of the IRA campaign on Twitter also found Black accounts to have the strongest relationship with engagement on that platform [16]. Here, we investigate whether this was the case for the Facebook and Instagram IRA campaigns, by examining whether IRA content concerning particular communities (e.g. the Black community, among others) and issues consistently received higher engagement.

On the topic of the IRA's social media interference campaign in the 2016 U.S. presidential election, the only previous studies that use a dataset as extensive as ours are Howard *et al.* [6] and DiResta *el al.* [7]. However, as noted above, these studies provide more preliminary findings and do not investigate the relationships between post features and user engagement. Several previous studies have examined the IRA campaign on Twitter, for which the data has been made public. Badawy *et al.* [14] focus on identifying features predictive of whether a Twitter user reshared at least one tweet from an IRA Twitter account. Bail *et al.* [15] used data from a one-month period in 2017 and a survey of Twitter users to estimate effects of engaging with IRA Twitter content on opinion, while highlighting their "inability to determine whether IRA accounts influenced the 2016 presidential election" (as indeed the public Twitter data, and the Facebook and Instagram SSCI data, do not warrant strong causal conclusions, as we will consider in the Discussion section). Freelon et *al.* [16] also studied the Twitter IRA campaign, investigating ideological and racial asymmetries in the numbers of IRA accounts, IRA posts, and engagement levels. Given that the Twitter IRA campaign has already been studied to a large extent, and that Twitter has a much smaller user base in the U.S. than Facebook and Instagram, with the latter two social media platforms having the largest userbase in the U.S. during the 2015-2017 period studied [17, 18], we focus on the Facebook and Instagram IRA campaigns.

Other previous studies have used publicly available data of IRA Facebook ads [19], or a combination of public Twitter data, Facebook ads data, and some photos and videos from some public IRA Facebook pages inferred using browser logs [20]. Therefore, the relationships between post features and engagement for organic (non-paid) Facebook and Instagram posts have not yet been studied. Indeed these organic posts were much more numerous than ads [6], however media and policy discussions this far have often focused on paid ads.

This study seeks to fill this research gap, focusing on the following research questions (RQs) regarding the IRA's Facebook and Instagram campaign:

**RQ 1** Did IRA content targeting particular communities and issues (e.g. Black communities, as has been suggested in media accounts and in empirical accounts of the IRA campaign on Twitter [16]) consistently get higher engagement?

**RQ 2** Besides communities targeted, what were the relationships of other textual and non-textual characteristics of the IRA's posts (e.g. sentiment, timing, use of hashtags, use of links or visual media) with engagement?

**RQ 3** How did these relationships differ across the two platforms?

Beyond the IRA campaign, previous studies of disinformation on social media around the 2016 U.S. election period has focused on examining social media posts that included links (URLs) to disinformation. This line of research has focused on analyzing Facebook [21] or Twitter [22, 23] posts that contained a link to a known external source of "fake" (or "junk") news, or on analysing Twitter posts linking to a known external fact-checking source [24]. Instead, our approach is to cover the online interference campaigns run by the IRA on Facebook and on Instagram in their entirety, rather than focusing on only those posts that contain links to specific external websites. Indeed, in the organic IRA campaign on Facebook and Instagram during 2015-2017, we find that the vast majority of posts did not contain a link (only 3.5% of Facebook posts did; this is not applicable to Instagram as links do not work in Instagram posts). Further still, we find a negative relationship between including a URL in a Facebook post and user engagement with that post. That is, not only were external links not frequent in the IRA's Facebook and Instagram activity, they were not even associated with high user engagement. Instead we measure relationships between engagement and a broader range of features of a whole online interference operation (of which links are only one element), across the two most popular social media platforms in the U.S. We cover more than two and a half years – before, during and after the 2016 U.S. election (January 2015 to August 2017 for Facebook and to October 2017 for Instagram), going beyond the few weeks or months prior to the 2016 election that have been the focus of several previous studies of disinformation and IRA online activity [14, 15, 21, 22].

**Methods**

*Data*

We use non-public data of Facebook and Instagram posts (N = 67,502 posts on Facebook, produced by 81 pages; N = 116,205 posts on Instagram, produced by 133 accounts), spanning the period 2015-2017. This is data that Facebook provided to the SSCI of organic social media posts uploaded on its Facebook and Instagram platforms by pages and accounts the company identified as being managed by IRA staff. The data was provided under a Non-Disclosure Agreement with the SSCI in order for analyses to be conducted in a rigorous and secure manner. To our knowledge, this is the most comprehensive dataset of organic IRA social media posts made externally available by Facebook.

The data was provided in tabular format, whereby each row (record) corresponded to a post, with columns for a post's ID, the date it was uploaded, name of Facebook page or Instagram account that posted it, text, number of likes, and number of comments. For Facebook posts, columns also included the number of shares, number of emoji reactions, link (URL) included in the post if any, and ID for media (photo or video) included in the post if any.

*Feature Extraction*

For each post, we extract three types of features: features about the audience that a post addresses, textual features, and non-textual features. We use these features to predict overall engagement with a post.

The audience feature is inferred based on the name and content of the Facebook page or Instagram account that uploaded the post, and is measured at two levels of granularity. First, it is measured at the audience level, taking four possible values: {Right, Black, Marginalised, Other}, where "Right" corresponds to posts from pages addressing right-wing audiences; "Black" for pages addressing African American audiences; "Marginalised" for pages addressing other marginalised groups (e.g. women, LGBT, Latin American, Muslim, or Native American groups); and "Other" for pages addressing all other audiences or no single specific audience (e.g. pages addressing liberal audiences, or communities local to a specific area, or pages posting memes and similar internet culture content). These four categories were agreed upon by all three authors unanimously, and they stem from the findings published in previous studies of these same datasets [6, 7].

Next, this audience variable is also measured at a finer granularity: for each of these four audience category values, at the sub-audience level, taking values such as Right – Second

Amendment; Black – Identity; Marginalised – LGBT. This sub-audience feature is also informed by the above two studies [6, 7]. The classification of the sub-audience addressed by a Facebook page or Instagram account was conducted by the first two authors, based on the name of the page or account and on a random sample of posts uploaded by that page. Supplementary Table 1 lists all audience and sub-audience values.

We note that, for the Facebook and Instagram IRA campaign data that we study, we group accounts targeting marginalised communities under marginalised audiences and sub-audiences, rather than under a 'liberal' or 'left' category, as was one in e.g. [14] for the Twitter IRA campaign, where accounts such as '@lgbtunitedcom', '@muslimsinusa' were categorised in a 'Left trolls' category rather than in a marginalised group category. This was done for several reasons. The key reason was that, on Facebook and Instagram, accounts that addressed marginalised groups and issues did not promote left or liberal viewpoints, but rather they mostly posted about not directly political issues specific to that group identity, and the minority of their posts that did relate to politics was about alienating these groups from the left, the liberals, and/or from Hillary Clinton, trying to instil distrust and apathy towards the election and encouraging people not to vote [6]. This was in contrast to the accounts we labelled as right-related, which were labelled thus because they did actively promote right-wing views and encouraged people to vote for Donald Trump [6]. Furthermore, there was a diverse set of marginalised groups being addressed by the IRA (for example: women and the women's movement, targeted by e.g. the @feminism_tag Instagram account; LGBT communities, targeted by e.g. the 'LGBT United' Facebook page; Muslim Americans targeted by e.g. the 'United Muslims of America' Facebook page; Latin Americans targeted by e.g. the 'Brown Power' Facebook page). There were one or multiple accounts and pages dedicated to each of these marginalised groups and posting about that specific group's issues and identity, rather than having catch-all liberal-targeting accounts and pages dedicated to posting liberal or pro-left content while sometimes mentioning various marginalised groups as part of an overall pro-left agenda.

Beyond audiences and sub-audiences, the textual features we that measure include the length in characters of the text contained in a post, and a separate sentiment measure of a post's text, representing an aggregate score of how positively and how negatively charged all the words in the post are, as inferred by the VADER sentiment analysis model for social media [32]. Non-textual features capture the temporal dimension (the date a post was uploaded), as well as the use of media, links, and hashtags, i.e. whether a link was included in a Facebook post; whether an image or video was included in a Facebook post; the number of hashtags included in an Instagram post.

We use post engagement as the outcome variable, measured as the sum of all types of user engagement a post has received, i.e. the total count of likes, comments, shares, and emoji reactions for a Facebook post, and total number of likes and comments of an Instagram post.

In more detail, we extracted and measured the following features (variables), for each Facebook and/or Instagram post:

- Engagement: The sum of all types of user engagement a post has received, i.e. the sum of Likes and Comments for Instagram, and on Facebook this also includes the sum of Shares and Emoji reactions. On Facebook, there are five emoji reactions (Love, Haha, Wow, Angry, Sad), however the dataset does not contain values for each one, but instead gives the total sum of all five together.
- Link: Equals one if the Facebook post link field is non-empty, i.e. if the post contains a URL link, and zero otherwise.
- Photo or video: Equals one if the Facebook post 'photo or video' field is non-empty, i.e. if the post contains a URL link, and zero otherwise.
- Number of hashtags: A count of the number of hashtags in an Instagram post.
- Date: The date a post was posted onto Facebook or Instagram, offset from 01/01/2015; i.e. how many days after 01/01/2015 a post was posted.
- Text length: The number of characters in the text field of a post.
- Sentiment: The VADER compound score [32] of the text field of the post. We use VADER as it is a state-of-the-art dictionary-based sentiment measure developed specifically for social media posts, that accounts for the sentiment intensity (i.e. strength of positivity or negativity) of each word in the text. This score can be thought of as an aggregate measure of how strongly positive or strongly negative the overall wording of a post's text is. [Due to the size of our dataset (hundreds of thousands of posts) and the data confidentiality constraints, it was not possible to obtain the human annotations of post sentiment required for supervised machine learning-based sentiment analysis, so a dictionary-based sentiment analysis approach was the most appropriate in our case.]

- Audience and sub-audience: A given post is assigned the audience and sub-audience class of the Facebook page or Instagram account that posted it. That is, these features are measured at the level of the page or account, as discussed above.

*Predictive modelling*

In our predictive models, we use engagement as the dependent variable, and all the other features mentioned above as the independent variables. Since our dependent variable is a count of engagements, and we are interested in measuring its relationships with the independent variables, we use regression modelling. Previous studies of misinformation or of the IRA campaign on social media around the 2016 U.S. election have used either linear regression models (e.g. [14], [21]), or nonlinear decision tree ensemble machine learning models (e.g. [15], where the question is modelled as a classification task with a binary outcome variable).

Here, instead of using only linear regression models, or only nonlinear decision tree ensemble machine learning models, we use both, in order to also establish which relationships persist across models types, regardless of the functional form of the models and of their bias-variance trade-off characteristics.

For the linear models, we report here the results of the Negative Binomial model (the most appropriate model for our positively skewed count data, e.g. per Cameron and Trivedi [33]), with details on the implementation and results of these and of three other linear models (OLS, OLS on Box-Cox log-transformed dependent variable, Poisson) in the Supplementary Section S.2.

For the nonlinear machine learning models, we use decision-tree ensemble regressors, specifically Random Forests and Gradient Boosting, and summarize the results of the model with the best fit here, reporting their goodness of fit measures and full results in the Supplementary Section S.3.

**Results**

Our unique data and extensive quantitative modelling allow us to provide the first analysis of the relationships of IRA organic post characteristics with user engagement, and compare these between Facebook and Instagram.

*Descriptive statistics of post volume and engagement on Facebook and Instagram*

We find some similarities and differences in the volume of posts that the IRA uploaded to Facebook versus those uploaded to Instagram, as shown in Table 1. Table 1 shows that, in total, the IRA uploaded many more posts to Instagram than on Facebook, and received much more engagement on Instagram than on Facebook. On a per-category basis, a similar volume of posts was uploaded to Facebook and to Instagram for the Other category (N is in the seven thousands on both cases), and for Marginalised (in the 11-12 thousands). However, the IRA's posting strategy was different on Facebook versus Instagram for the Black and the Right-wing audiences: many more Black and Right-related IRA posts were uploaded to Instagram than on Facebook (indeed more than twice as many Black-related posts on Instagram than on Facebook).

|  | **Facebook** | | **Instagram** | |
| --- | --- | --- | --- | --- |
|  | **N** | **Engagement** | **N** | **Engagement** |
| Black | 23,306 | 18,167,564 | 54,983 | 75,938,997 |
| Marginalised (other than Black) | 11,566 | 11,534,022 | 12,895 | 32,011,478 |
| Right | 25,500 | 48,476,504 | 40,348 | 77,043,768 |
| Other (none of the above) | 7,130 | 462,352 | 7,951 | 3,638,990 |
| *Total* | *67,502* | *78,640,442* | *116,205* | *188,635,215* |

**Table 1**. **Audience categories for Facebook and Instagram**, along with the respective count (N) of posts in each and total engagement for these posts.

In terms of per-post engagement on Facebook and Instagram, some IRA posts received hundreds of thousands of user engagements, with the maximum observed engagement on Facebook nearing one million (986,259 engagements, per Table 2). However, posts with hundreds of thousands of engagements were extremely rare: on both platforms, 99% of IRA posts received fewer than 20,000 engagements. Indeed, engagement is highly skewed to the right, as shown in Fig. 1, with most posts getting no engagements (as is common for attention and engagement on social media [25, 26, 27, 28], including engagement with the IRA's social media posts [21]). Per Table 2, half the posts on both platforms received engagement in the hundreds. The least engaged-with quarter of the Facebook and Instagram posts only received up to twenty-one and 152 engagements respectively (25$^{th}$ percentile in Table 2). The most-engaged with quarter of Facebook and Instagram posts received at least one thousand engagements (75$^{th}$ percentile in Table 2). The top 1% most engaged-with posts (which corresponds to hundreds of Facebook posts and thousands of Instagram posts, see N in Table 2) each received more than 10,000 engagements on Facebook and more than 15,000 engagements on Instagram (99$^{th}$ percentile in Table 2). So even though half the posts did not receive thousands of engagements, because the total number of IRA posts (N) is so large, there are still several thousands of IRA posts that each received thousands or even tens of thousands of engagements.

|  | Min | Max | Mean | Standard Deviation | Percentiles | | | | | |
|---|---|---|---|---|---|---|---|---|---|---|
|  |  |  |  |  | 25$^{th}$ | 50$^{th}$ | 75$^{th}$ | 90$^{th}$ | 95$^{th}$ | 99$^{th}$ |
| **Facebook** | 0 | 986,259 | 1,165 | 7,598 | 21 | 229 | 1,069 | 2,753 | 4,361 | 10,541 |
| **Instagram** | 0 | 260,913 | 1,623 | 3,559 | 152 | 574 | 1,712 | 3,947 | 6,109 | 15,957 |

**Table 2. Descriptive statistics of per-post engagement for Facebook and Instagram.**

Comparing Instagram and Facebook in Table 2, even if the maximum per-post engagement value for Instagram is lower, the mean and all percentiles for Instagram are higher than for Facebook, so on Instagram engagement levels are higher on average. Facebook data is twice as variable as Instagram's (twice as large standard deviation, Table 2), hence levels of engagement on Facebook are harder to predict than for Instagram, as we shall see.

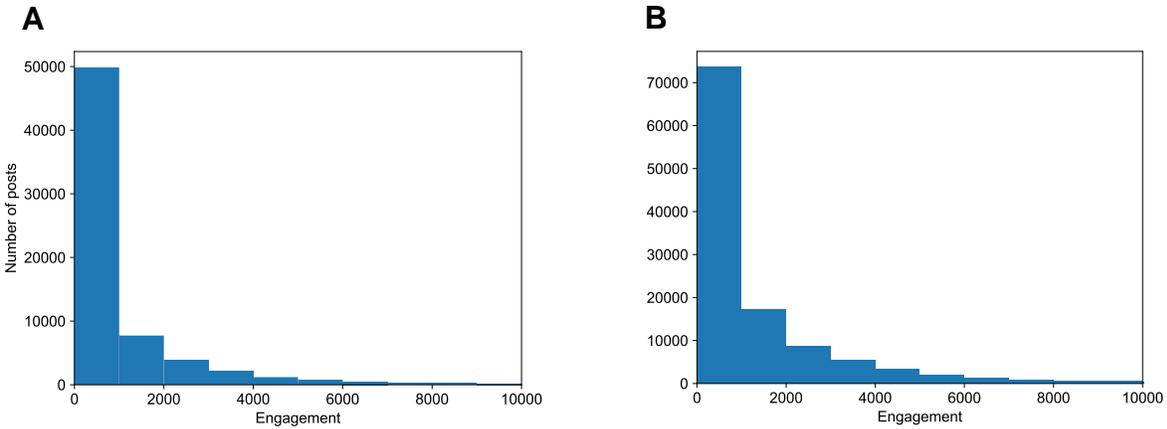

**Fig. 1**. **Histogram of engagement on Facebook (A) and Instagram (B).** For better visibility, the x-axis is truncated at 10000 engagements.

Plotting the distribution of engagement (excluding outliers from the visualisation) versus audience (Fig. 2), we see that for both platforms engagement (including mean and median values) tends to be higher for posts addressing right-wing and marginalised audiences than for posts addressing Black audiences, and in turn engagement for these three audiences tends to be higher than for other audiences.

We next use regression models to investigate whether this relationship holds when we control for further independent variables, and we also examine the relationship of these additional variables with engagement.

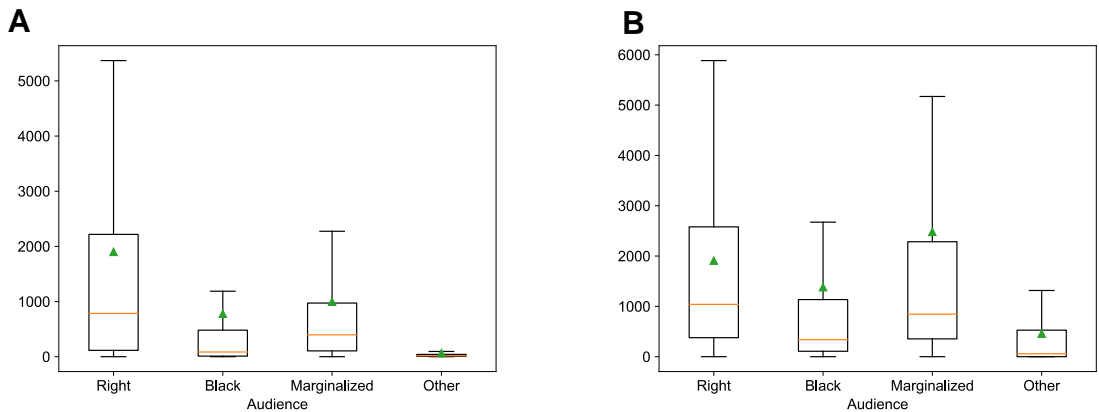

**Fig. 2. Box-and-whiskers plots of engagement grouped by audience, for Facebook (A) and Instagram (B).** The triangle indicates the mean.

*Predictors of engagement on Facebook and Instagram, using the audience variable*

We regress engagement on the audience, date, text length, and sentiment variables (for both platforms), and for Facebook also on the link and the photo or video variables, while for Instagram also on the number of hashtags variable. Table 3 shows that, on both platforms, posts targeting right-wing audiences have a stronger positive relationship with engagement than those targeting marginalised audiences, followed by those targeting Black audiences. Indeed, effect sizes for posts targeting right-wing audiences are far greater than for those targeting Black audiences, differing by a factor of 2.61 for Facebook ($e^{(3.53-2.57)} = 2.61$), and by a factor of 1.70 for Instagram ($e^{(1.73-1.20)} = 1.70$). Effect sizes for non-Black marginalised audiences are 1.26 times higher than those for Black audiences on Facebook ($e^{(2.80-2.57)} = 1.26$), and 1.66 times higher on Instagram ($e^{(1.70-1.20)} = 1.66$).

|                              | Facebook      | Instagram     |
| ---------------------------- | ------------- | ------------- |
| Right                        | 3.53***       | 1.73***       |
|                              | (0.014)       | (0.01)        |
| Marginalised (other than-Black) | 2.80***    | 1.70***       |
|                              | (0.016)       | (0.013)       |
| Black                        | 2.57***       | 1.20***       |
|                              | (0.014)       | (0.01)        |
| Photo or video               | 0.41***       |               |
|                              | (0.023)       |               |
| Date                         | 0.001***      | 0.004***      |
|                              | (0.00002)     | (0.00001)     |
| Text length                  | $-0.001$***   | $-0.001$***   |
|                              | (0.00001)     | (0.00001)     |
| Sentiment                    | $-0.18$***    | 0.02***       |
|                              | (0.007)       | (0.005)       |
| Link                         | $-1.52$***    |               |
|                              | (0.02)        |               |
| Number of hashtags           |               | 0.007***      |
|                              |               | (0)           |
| Intercept                    | 3.42***       | 2.88***       |
|                              | (0.032)       | (0.012)       |
| N                            | 67502         | 116205        |

*$p < 0.1$; **$p < 0.05$; ***$p < 0.01$

**Table 3. Determinants of engagement with IRA posts on Facebook and Instagram, using the audience variable**. Negative Binomial regression models on all variables, showing results for the audience variables (full results in the Supplementary Section S.2). The dependent variable is the count of total engagement numbers a post has received. The reference category for audience is "Other." Standard errors in parentheses.

***Predictors of engagement on Facebook and Instagram, using the sub-audience variable***

Tables 4 and 5 again show regression results using all variables, separately for Facebook and for Instagram respectively, this time using the finer-grained audience sub-categories variable instead of the audience variable.

| Feature | Engagement | Feature | Engagement |
|---|---|---|---|
| Right – Nativist | 7.52*** | Marginalised – Native American | 3.51*** |
|  | (0.12) |  | (0.12) |
| Marginalised – Latin American | 7.52*** | Other – Internet Culture | 3.26*** |
|  | (0.12) |  | (0.12) |
| Right – Southern Identity | 7.23*** | Black – Self-Defense | 2.68*** |
|  | (0.12) |  | (0.12) |
| Right – Christian | 7.05*** | Right – Anti-Clinton | 0.69*** |
|  | (0.12) |  | (0.15) |
| Right – Patriotic | 6.90*** | Right – Internet Culture & Alt-Right | − 0.22 |
|  | (0.12) |  | (0.15) |
| Black – Social Justice & Activism | 6.75*** | Black – Religion | − 0.76*** |
|  | (0.12) |  | (0.14) |
| Right – Second Amendment | 6.65*** | Black – Other | − 2.26*** |
|  | (0.12) |  | (0.46) |
| Marginalised – Muslim | 6.28*** | Date | 0.0003*** |
|  | (0.12) |  | (0.00002) |
| Marginalised – LGBT | 5.97*** | Text length | − 0.0008*** |
|  | (0.12) |  | (0.00001) |
| Right – Veterans & Police | 5.71*** | Sentiment | − 0.0486*** |
|  | (0.12) |  | (0.007) |
| Right – Pro-Trump | 5.35*** | Photo or video | − 0.16*** |
|  | (0.37) |  | (0.02) |
| Black – Identity | 4.57*** | Link | − 1.39*** |
|  | (0.12) |  | (0.02) |
| Other – Liberal | 4.05*** | Intercept | 0.70*** |
|  | (0.12) |  | (0.12) |
| N | 67502 |  |  |

*$p < 0.1$; **$p < 0.05$; ***$p < 0.01$

**Table 4: Facebook - Determinants of engagement with IRA posts, using audience subcategories.**
Negative Binomial regression models on all variables, showing results for the audience subcategory variables (full results in the Supplementary Section S.2). The dependent variable is the count of total engagement numbers a post has received. The reference category for the audience subcategories is "Other – Unknown & Various". Standard errors in parentheses.

| Feature | Engagement | Feature | Engagement |
|---|---|---|---|
| Marginalised – LGBT | 2.62*** | Black – Social Justice & Activism | 1.16*** |
| | (0.02) | | (0.01) |
| Marginalised – Women | 2.54*** | Right – Second Amendment | 0.91*** |
| | (0.02) | | (0.03) |
| Right – Veterans & Police | 2.49*** | Marginalised – Native American | 0.43*** |
| | (0.01) | | (0.03) |
| Right – Patriotic | 2.21*** | Black – Self-Defense | – 0.16*** |
| | (0.01) | | (0.03) |
| Right – Christian | 2.11*** | Black – Other | – 0.22 |
| | (0.02) | | (0.28) |
| Black – Identity | 1.71*** | Other – Local interest | – 1.25*** |
| | (0.01) | | (0.08) |
| Right – Internet Culture & Alt- Right | 1.54*** | Number of hashtags | 0.008*** |
| | (0.02) | | (0) |
| Marginalised – Latin American | 1.53*** | Date | 0.004*** |
| | (0.02) | | (0.00001) |
| Right – Southern Identity | 1.32*** | Text length | – 0.00007*** |
| | (0.02) | | (0.000009) |
| Marginalised – Muslim | 1.31*** | Sentiment | – 0.05*** |
| | (0.02) | | (0.01) |
| Right – Nativist | 1.26*** | Intercept | 2.65*** |
| | (0.02) | | (0.01) |
| Other – Liberal | 1.19*** | | |
| | (0.02) | | |
| N | 116205 | | |

*$p < 0.1$; **$p < 0.05$; ***$p < 0.01$

**Table 5: Instagram - Determinants of engagement with IRA posts, using audiences subcategories.** Negative Binomial regression models on all variables, showing results for the audience subcategory variables (full results in the Supplementary Section S.2). The dependent variable is the count of total engagement numbers a post has received. The reference category for the audience subcategories is "Other – Unknown & Various". Standard errors in parentheses.

The following results from Tables 4 and 5 are consistent with results from the respective nonlinear machine learning model (Supplementary Section S.3): for the audience subcategories, on Facebook the Marginalised – Latin American and the Right – Nativist subcategories have consistently the top two most positive relationships with engagement, with Right – Southern Identity and Right – Christian completing the top four. The subcategories Black – Social Justice & Activism, Right – Patriotic, and Right – Second Amendment are also consistently in the top seven most positive relationships with engagement. Marginalised – LGBT and Marginalised – Muslim are also consistently in the top ten. Overall, we note that for Facebook the only Black sub-audience consistently in the top ten most positive relationships with engagement is Social Justice & Activism, and the rest of the top subcategories are generally from the right and marginalised audience categories. Black – Other has a consistently non-positive (i.e. negative or zero) relationship with engagement on Facebook, while Black – Religion and Right – Internet Culture & Alt-Right have a consistently negative relationship.

For Instagram, we find similarities and differences in the mix of sub-audiences having very positive relationships with engagement compared to Facebook. On Instagram, it is the Marginalised – Women and the Marginalised – LGBT sub-audiences that consistently have the top two most positive relationships with engagement, with Right – Veterans & Police having the third most positive relationship. Right – Christian, Black – Identity and Right – Patriotic have the fourth to sixth most positive relationships respectively. Similarly to Facebook, the most positive relationships come from the marginalised and right-wing audiences, with only one Black sub-audience in the top positive relationships; however, that sub-audience for Instagram is about celebrating Black identity, rather than about social justice and activism. Also, on Instagram, addressing the Marginalised – Women, Marginalised – LGBT or Right – Veterans & Police sub-audiences has a relatively much more positive relationship with engagement than for Facebook, while addressing groups like Marginalised – Latin American or Right – Nativist is not in the top few most positive relationships for Instagram in contrast to Facebook. Content on Instagram targeting the Right – Internet Culture & Alt-Right has a positive relationship with engagement, whereas it has a negative relationship on Facebook. On Facebook, Right – Southern Identity consistently has the third most positive relationships with engagement, however on Instagram it has a much weaker relationship to engagement, at ninth position in Table 5, and even lower, with zero relationship to engagement, in the nonlinear model ((Supplementary Section S.3).

There are some notable differences across the two platforms regarding which of the sub-audiences were associated with most and least engagement. Content related to the women's movement was in the top two most positive relationships with engagement for Instagram, while surprisingly on Facebook there were no pages dedicated to the women's movement. Content about Black – Social Justice & Activism issues had a much stronger positive association with engagement on Facebook than on Instagram, where instead Black – Identity was the subcategory of posts targeting the Black community which had the most positive relationship with engagement. Similarly, Right – Southern Identity had a much more positive relationship with engagement on Facebook than on Instagram.

Beyond the sub-audiences variables, the following relationships are observed for the textual and non-textual variables. For both Facebook and Instagram, the post date has a small positive relationship with engagement, meaning that posts uploaded later during the 2015-2017 interval received slightly more engagement on average. The nonlinear models show how this relationship is monotonically increasing for Instagram with a drop from July 2017 onward (Supplementary Fig. 3 A), while for Facebook there is a drop earlier, from March 2017 onward (Supplementary Fig. 2 A). This might be a reflection of Facebook and Instagram starting to ban IRA pages in 2017.

For text length, the linear negative binomial models find a small negative overall relationship with engagement for both Instagram and Facebook, while the nonlinear model highlights the nonlinearity in those relationships (Supplementary Figs. 2 B, 3 C) – for both Facebook and Instagram, an early positive relationship (peaking at around 163 and 50 characters respectively) is followed by a negative relationship for texts longer than around 345 characters for Facebook and 130 for Instagram. This indicates that IRA posts with shorter and snappier texts have a more positive relationship with engagement than longer, more rambling texts. This holds for both platforms, manifesting more acutely for Instagram.

Post sentiment for Facebook has a consistently negative relationship with engagement, with the nonlinear model (Supplementary Fig. 2 C) revealing a strongly positive relationship for extremely negative sentiment values (for sentiment values in the range -1.00 to -0.98), small positive relationships for narrow ranges of very negative sentiment values (-0.97 to -0.87) and very positive sentiment values (0.71 to 0.76), and very small-magnitude negative relationships everywhere else. For Instagram the linear model also finds an overall small negative relationship (Table 5), however the patterns detected by the nonlinear models (Supplementary Fig. 3 D) differ from those for Facebook, as there is a small negative relationship here everywhere except in the positive intervals of sentiment values of 0.41-0.53 (very small positive relationship) and 0.54-0.96 (stronger positive relationship). This indicates that, for Instagram, a positive relationship of sentiment with engagement exists only for highly positive sentiment (whereas for Facebook this also happens for the extremely negative sentiment). For Instagram we do not see a consistent relationship emerging between sentiment and engagement.

For Facebook, including a link in a post has a consistently negative relationship with engagement. The evidence for the relationship of including a photo or video with engagement is inconclusive, as the negative binomial linear model (Table 4) finds a negative relationship, while the nonlinear model finds a positive one.

For Instagram, the relationship of the number of hashtags used with engagement is found to be weakly positive by the negative binomial model (Table 5). The nonlinear model (Supplementary Fig. 3 B) shows a positive relationship for 1-5 hashtags (very strongly positive for 1-4 hashtags, with a peak at 2 hashtags) and a weak positive relationship for 28 or more hashtags, but it is weakly negative everywhere else, with a strongly negative relationship at 0 hashtags.

**Discussion**

This study uses a unique dataset and contributes several findings on the IRA's social media campaigns around the 2016 U.S. presidential election, helping anticipate engagement patterns with Russian social media content in the upcoming 2020 U.S. election.

We find that the most engaging content was not just, or even mainly, about Black issues. Rather, the content with highest engagement on average was about right-wing issues and issues related to non-Black marginalised groups. This is in contrast to popular narratives in the media, as well as to the findings regarding the IRA Twitter campaign [14]. Indeed, our findings on the Facebook and Instagram IRA campaign differ to those from the Twitter campaign (in [14]) in several regards, highlighting that there were differences in how various audiences engaged with IRA posts across different platforms, as well as in the IRA's posting strategy across the platforms. Most importantly, we find that on Facebook and Instagram the strongest predictor of engagement was not addressing the Black community, but rather addressing right-wing and other marginalised communities were the strongest predictors of engagement (Tables 3-5). Furthermore, while it has been found that, on Twitter, Black-presenting IRA accounts received more engagement than other identity categories[14], we find that this was not the case for Facebook and Instagram, where it was posts addressing right-wing and non-Black marginalised audiences that had a higher mean and median engagement (Fig. 2). In addition, while for Twitter it was found that there were 'substantially more' conservative-presenting tweets [14], this was not the case for Instagram (Table 1) where Black-related posts were the most numerous, nor quite for Facebook where there were only slightly more right-related posts (25,500) than Black-related ones (23,306).

The association that we find between right-wing IRA content and high user engagement parallels findings from other studies (albeit ones of different research questions, data and contexts) that, in the months prior to the 2016 election, more conservative-leaning social media users were more likely to share IRA content on Twitter [14], and to engage with known sources of fake (or "junk") news on Facebook [21] and on Twitter [22].

It is worth highlighting that the association that we find between IRA content targeting marginalised communities and high engagement is notable, not least because, beyond the Black community, other marginalised communities were generally not studied in previous analyses of audience engagement with the IRA's online campaigns.

In more detail, regarding specific sub-audiences and issues, we find that IRA campaign attempted to mobilize right-wing audiences around nativist and patriotic themes, narratives about veterans and the police, and contentious issues such as Christianity, the South, and the Second Amendment. Regarding marginalised communities, the most engaged-with content related to the women's movement, the LGBT movement, Latin American and Muslim communities. Many posts were not directly political or election-related. Those that were directly political and addressed right-wing audiences tended to include messages in favour of the Trump campaign, while those addressing marginalised communities tended to attempt to discredit Hilary Clinton as a candidate that could help these communities, and to encourage apathy and non-participation in the election.

Overall, the most engaged-with content related to divisive issues across the board, targeting many different communities. These findings have important implications for how to defend against future foreign online influence operations around elections, indicating that solutions should not focus only on one political or social group, but should rather address divisions across a range of socio-political issues and groups. Many fissures in U.S. public life were taken advantage of and used effectively to engage social media users. In order to prevent these being exploited again by foreign actors in the future, it is crucial to ramp up efforts to reduce polarization and tribalism, promote constructive debate around contentious issues, and engage, integrate and promote rights of marginalised communities.

Platform solutions should allow all users to check the background information of social media pages and accounts. To this end, platforms should display relevant information to users, such as when a page was first launched, any previous names it had, who is behind the page, the ads bought by the page or account or its administrators, including what groups these ads targeted, when, in what currency they were purchased, and from where in the world each post is uploaded. Platforms have already taken some of these steps, and our findings reinforce the importance of expanding and expediting this effort and making this information clear and accessible to all.

Additionally, there are some differences across the two platforms, regarding which of the various sub-communities were associated with most and least engagement (e.g. for the sub-audiences of Marginalised – Women, Black – Identity versus Black – Social Justice and Activism, and Right – Southern Identity, as discussed in the previous section). These patterns raise promising questions for future research, such as whether some social groups are more active on one platform than others; or whether identity plays a stronger role on Instagram than on Facebook.

Furthermore, we find that shorter messages were associated with higher engagement on Facebook and Instagram, while sentiment has an overall weak negative relationship with engagement on Facebook (Dutt *et al.* [19] also found a negative relationship between sentiment and engagement for IRA ads on Facebook). On Instagram, using more than a few hashtags (which can supposedly boost engagement by exposing a post to more people than just the account's followers) is, surprisingly, associated with lower engagement with the IRA's posts.

Our models include several different kinds of possible causes of user engagement, many of which constitute possible confounders, i.e. common causes behind user exposure to IRA content and user engagement with IRA content. The regression coefficients presented above can be interpreted as causal effects of the independent variables on user engagement only insofar as one is willing to assume that there is no unobserved confounding from other causes not included in this data. However, such an assumption has limitations, as it is difficult to ignore potential unobserved confounders, such as the audiences' pre-existing attitudes and interests, the popularity or newsworthiness of the messages and topics being discussed in the IRA posts, the Facebook and Instagram feed algorithms, or the possibility of other IRA or non-IRA groups on Facebook or on other platforms coordinating in order to promote certain IRA posts and make them accumulate lots of engagement quickly. The possible existence of such confounders means we cannot robustly attribute user engagement with this content to the IRA alone [29, 30, 31]. In addition, the data does not include posts from any non-IRA pages and accounts of a similar nature (e.g. domestic pages posting a similar mix of not directly political content and inflammatory content around the same divisive issues), which would be useful for comparing and contextualizing engagement with IRA content versus engagement with similar but non-IRA content.

Furthermore, the confidential manner in which this data was released imposed the limitation that it was not possible to e.g. obtain post-level labels for sentiment to be used by supervised machine learning models, or to extract linguistic, rhetoric, framing, and narrative tropes used by the IRA (e.g. us-versus-them narratives). This limited our focus to those variables that could be automatically extracted, with a minimal amount of human annotation.

That is, in terms of conducting a complete causal analysis, this study has gone as far as the limitations of this data would allow. We note that there are limits to the plausibility of the no-unmeasured-confounding assumption, so it is not possible to make unqualified causal claims about the causes of on-platform engagement based on this data, let alone about the causal effect of the IRA content on voting behaviour and the election outcome. Based on this data, one can make associational claims, and at best qualified causal claims under a no-confounding assumption, about the effects of the IRA's posts on online user engagement (with engagement being a valuable metric of how much people were exposed to, paid attention to, and interacted with IRA content). Hence, we interpret our findings in associational terms, and we uncover strong and important relationships across a range of key causal variables, as well as demonstrating how these differ across the two platforms. Our findings highlight a need and an opportunity for more detailed data of the kinds discussed above to be made available, to enable future research into the possible causal pathways behind the associational relationships we have identified, and our findings highlight important features of the IRA campaign the study of which can be prioritized in future work.

We find significant and interesting relationships between audiences, sub-audiences, textual and non-textual features, and engagement, with similarities and differences across the two platforms. We establish that it was targeting right-wing and non-Black marginalised communities that was associated with the most engagement, and that a wide range of sub-audiences and divisive issues were associated with high engagement, with similarities and differences on Facebook versus Instagram. Our findings contribute towards policy and research, by highlighting the importance of combating socio-political divisions across the board rather than focusing on one social or political group only, and of platforms making page information transparent to all users. We emphasize the importance of platforms making available richer data (while respecting user privacy), including measurements of confounding causal factors, in order to get closer to understanding the causal relationships between the IRA content's features, the broad characteristics of the communities who actually engaged with this content, and ultimately the effect of IRA posts on political attitudes and election behaviour.

**References**

[1] J. Abbruzzese, "Mark Zuckerberg: Facebook caught Russia and Iran trying to interfere in 2020," 2019.

[2] D. Chiacu, "FBI Director Wray: Russia intent on interfering with U.S. elections," 2019.

[3] A. Sebenius, "Russian Internet Trolls Are Apparently Switching Strategies for 2020 U.S. Elections," *Time,* 2019.

[4] S. Jack, "Facebook's Zuckerberg defends actions on virus misinformation," *BBC News,* 2020.

[5] M. Zuckerberg, "Mark Zuckerberg: Protecting democracy is an arms race. Here's how Facebook can help," *The Washington Post,* 2018.

[6] P. N. Howard, B. Ganesh, D. Liotsiou, J. Kelly and C. François, "The IRA, Social Media and Political Polarization in the United States, 2012-2018," Working Paper, Project on Computational Propaganda, Oxford, 2018.

[7] R. DiResta, K. Shaffer, B. Ruppel, D. Sullivan, R. Matney, R. Fox, J. Albright and B. Johnson, "The Tactics & Tropes of the Internet Research Agency," New Knowledge, 2018.

[8] K. Dilanian and B. Popken, "Russia favored Trump, targeted African-Americans with election meddling, reports say," *NBC News,* 2018.

[9] A. C. Madrigal, "When is a meme a foreign-influence operation?," *The Atlantic,* 2018.

[10] C. Riotta, "Russia targeted Black voters in attempts to suppress Democrat turnout in presidential election, Senate report says," *The Independent,* 2018.

[11] S. Shane and S. Frenkel, "Russian 2016 Influence Operation Targeted African-Americans on Social Media," *The New York Times,* 2018.

[12] J. Swaine, "Russian propagandists targeted African Americans to influence 2016 US election," *The Guardian,* 2018.

[13] D. Volz, "Russians Took Aim at Black Voters to Boost Trump, Reports to Senate Find," *The Wall Street Journal,* 2018.

[14] A. Badawy, K. Lerman and E. Ferrara, "Who falls for online political manipulation?," in *Companion Proceedings of The 2019 World Wide Web Conference*, 2019.

[15] C. A. Bail, B. Guay, E. Maloney, A. Combs, D. S. Hillygus, F. Merhout, D. Freelon and A. Volfovsky, "Assessing the Russian Internet Research Agency's impact on the political attitudes and behaviors of American Twitter users in late 2017.," *Proceedings of the National Academy of Sciences,* no. 117, pp. 243-250, 2019.

[16] D. Freelon, M. Bossetta, C. Wells, J. Lukito, Y. Xia and K. Adams, "Black Trolls Matter: Racial and Ideological Asymmetries in Social Media Disinformation," *Social Science Computer Review,* 2020.

[17] S. Greenwood, A. Perrin and M. Duggan, "Social Media Update 2016," Pew Research Center,, 2016.

[18] "Social Media Fact Sheet," Pew Research Center, 2019.

[19] R. Dutt, A. Deb and E. Ferrara, ""Senator, We Sell Ads": Analysis of the 2016 Russian Facebook Ads Campaign.," in *International conference on intelligent information technologies*, 2018.

[20] A. Spangher, G. Ranade, B. Nushi, A. Fourney and E. Horvitz, "Analysis of Strategy and Spread of Russia-sponsored Content in the US in 2017," arXiv preprint arXiv:1810.10033, 2018.

[21] A. Guess, J. Nagler and J. Tucker, "Less than you think: Prevalence and predictors of fake news dissemination on Facebook.," *Science Advances,* 2019.

[22] N. Grinberg, K. Joseph, L. Friedland, B. Swire-Thompson and D. Lazer, "Fake news on Twitter during the 2016 U.S. presidential election," *Science,* no. 363, p. 374–378, 2019.

[23] A. Bovet and H. A. Makse, "Influence of fake news in Twitter during the 2016 US presidential election.," *2019,* no. 10, p. 1–14 , Nature Communications.

[24] S. Vosoughi, D. Roy and S. Aral, "The spread of true and false news online," *Science,* vol. 359, p. 1146–1151, 2018.

[25] M. J. M and E. G. Altmann, "Predictability of extreme events in social media.," *PLoS ONE 9(11): e111506,* 2014.

[26] M. E. J. Newman, "Power laws, Pareto distributions and Zipf's law," *Contemporary physics ,* vol. 46, no. 5, pp. 323-351, 2005.

[27] E. Bakshy, J. M. Hofman, W. A. Mason and W. D. J, "Everyone's an influencer: quantifying influence on Twitter," in *Proceedings of the fourth ACM international conference on Web search and data mining*, 2011.

[28] J. Cheng, L. Adamic, P. A. Dow, J. M. Kleinberg and J. Leskovec, "Can cascades be predicted?," in *Proceedings of the 23rd international conference on World wide web, WWW'14*, 2014.

[29] D. Liotsiou, L. Moreau and S. Halford, "Social influence: From contagion to a richer causal understanding," in *International Conference on Social Informatics*, 2016.

[30] D. Liotsiou and P. N. Howard, "Measuring the Influence of Online Misinformation: A Hierarchy of Social Media Data," in *The 5th International Conference on Computational Social Science (IC2S2)*, 2019.

[31] S. Aral and D. Eckles, "Protecting elections from social media manipulation," *Science,* no. 365, p. 858–861, 2019.

[32] C. J. Hutto and E. Gilbert, "Vader: A parsimonious rule-based model for sentiment analysis of social media text," in *Proceedings of the 8th International AAAI Conference on Weblogs and Social Media (ICWSM)*, 2014.

[33] A. C. Cameron and P. K. Trivedi, Regression Analysis of Count Data, Cambridge: Cambridge University Press, 2013.

**Acknowledgments**

We thank the Senate Select Intelligence Committee (SSCI) for access to data. The research was approved by the University of Oxford's Research Ethics Committee, CUREC OII C1A 15 044, C1A 17 054. Any opinions, findings, and conclusions or recommendations expressed in this material are those of the researchers and do not necessarily reflect the views of the funders, or the University of Oxford .

**Funding**

This work is funded by the European Research Council through the grant "Computational Propaganda: Investigating the Impact of Algorithms and Bots on Political Discourse in Europe," Proposal 648311, 2015-2020, Philip N. Howard, Principal Investigator, and the grant "Restoring Trust in Social Media Civic Engagement," Proposal Number: 767454, 2017-2018, Philip N. Howard, Principal Investigator. We are grateful for additional support from the Adessium Foundation, Hewlett Foundation, Newmark Foundation, and Ford Foundation.

**Author contributions**

All authors conceived of the study. P.N.H. provided data. D.L. and B.G. processed and annotated the data. D.L. devised and carried out the statistical and machine learning modelling and regression analyses and produced visualizations. D.L. and B.G. performed the literature review. D.L. wrote the paper. D.L and P.N.H. revised the paper.

**Competing interests**

The authors declare no competing interests.

**Data and materials availability**

The data was provided to the authors by the U.S. Senate Select Committee on Intelligence (SSCI) under confidentiality restrictions, and access to it is determined by the SSCI.

# Supplementary Information

**Contents**

## S.1 Audience categories and sub-categories

|  | Facebook | | Instagram | |
|---|---:|---:|---:|---:|
|  | **N** | **Engagement** | **N** | **Engagement** |
| **Black** | **23306** | **18167564** | **54983** | **75938997** |
| – Identity | 9739 | 1724312 | 33249 | 64822303 |
| – Other | 29 | 6 | 14 | 384 |
| – Religion | 530 | 349 | - | - |
| – Self-Defense | 1477 | 26399 | 1870 | 411046 |
| – Social Justice & Activism | 11531 | 16416498 | 19850 | 10705264 |
| **Marginalised (other than Black)** | **11566** | **11534022** | **12895** | **32011478** |
| – Latin American | 1083 | 3902205 | 2301 | 4585594 |
| – LGBT | 5338 | 3388324 | 3288 | 10943604 |
| – Muslim | 4173 | 4189008 | 3485 | 2320777 |
| – Native American | 972 | 54485 | 1707 | 949387 |
| – Women | - | - | 2114 | 13212116 |
| **Right** | **25500** | **48476504** | **40348** | **77043768** |
| – Anti-Clinton | 170 | 628 | - | - |
| – Christian | 1654 | 3660002 | 3421 | 7383993 |
| – Internet Culture & Alt-Right | 206 | 330 | 2852 | 3834878 |
| – Nativist | 5228 | 14763062 | 5843 | 5390767 |
| – Patriotic | 8124 | 11673541 | 14284 | 25681556 |
| – Pro-Trump | 8 | 488 | - | - |
| – Second Amendment | 1159 | 1668574 | 1960 | 2733946 |
| – Southern Identity | 6470 | 15337910 | 4348 | 5650551 |
| – Veterans & Police | 2481 | 1371969 | 7668 | 26370059 |
| **Other (none of the above)** | **7130** | **462352** | **7951** | **3638990** |
| – Internet Culture | 4886 | 248763 | - | - |
| – Liberal | 2074 | 213490 | 2361 | 2899782 |
| – Local Issues | - | - | 401 | 2638 |
| – Unknown & Various | 170 | 99 | 5189 | 736570 |
| *Total* | **67502** | **78640442** | **116205** | **188635215** |

**Supplementary Table 1**. Audience categories and their sub-categories (sub-audiences), for Facebook and Instagram, along with the respective count (N) of posts in each and total engagement for these posts.

## S.2 Linear regression results for Facebook and Instagram

For the linear models, for all feature combinations across Facebook and Instagram, we follow the same sequence of models. We start with a simple and interpretable Ordinary Least Squares (OLS) regression model, however a Quartile-Quartile plot of the residuals reveals that the model's heteroscedasticity assumption is violated (Supplementary Figure 1 A for Facebook and C for Instagram). So we next use OLS with a Box-Cox natural-logarithm transformation of the dependent variable (ln(y+0.5)), which greatly improves heteroscedasticity as indicated in the QQ plot (Supplementary Figure 1 B for Facebook and D for Instagram). But since our independent variable represents count data where high values are rare and most values are near zero (Fig. 1 of main article), Poisson regression is appropriate (e.g. per Cameron and Trivedi, 2013). We implement Poisson regression but find that the equidispersion requirement is violated, as our data is overdispersed (see Supplementary Tables 3, 5, 8, 12, where the value for (Pearson $X^2$/(Df Residuals) is much greater than one), so to address this we use Negative Binomial, which greatly improves the overdispersion goodness of fit metric is still greater than one but much smaller than for the Poisson regression models (see Supplementary Tables 3, 5, 9, 13, (Pearson $X^2$/(Df Residuals) metric).

      We next report the results of all models, starting with the feature combination that includes audience variables, and proceeding to the feature combination that instead uses sub-audience variables, in each case reporting Facebook results first followed by Instagram results.

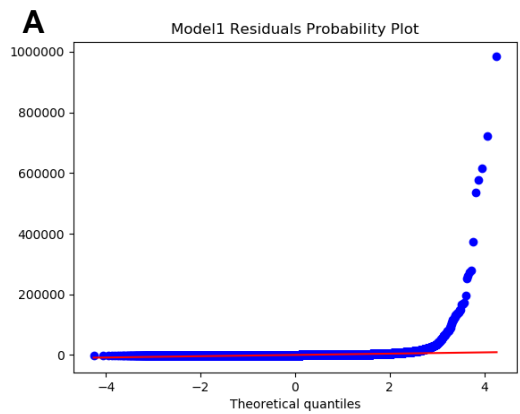
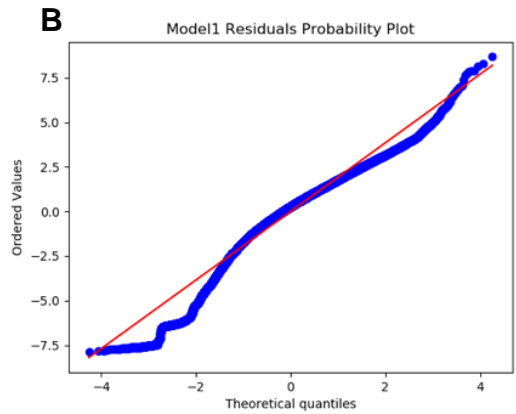
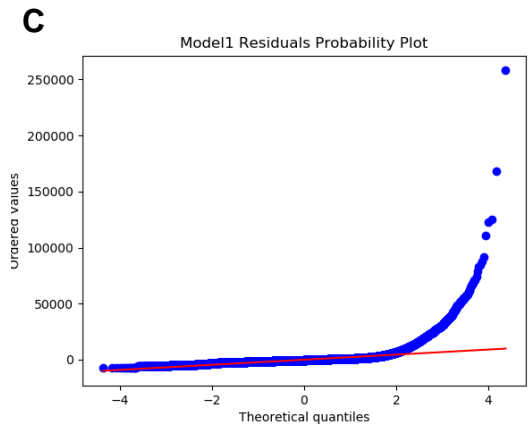
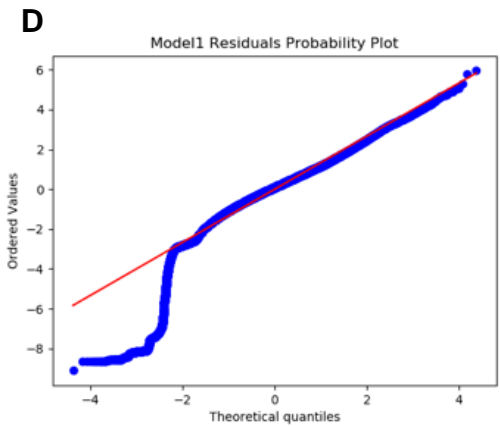

**Supplementary Figure 1.** Quartile-Quartile (QQ) plots for OLS regressions for Facebook (A, B) and Instagram (C, D), using the sub-audience variable, using engagement as the dependent variable (A, C) versus using the Box-Cox log-transformed engagement, i.e. ln(engagement + 0.5), as the dependent variable (B, D)

**Audiences**

Facebook

|  | **Engagement** |  | **Ln(Engagement+0.5)** |
|---|---|---|---|
| Right | 2004.05*** | Right | 3.09*** |
|  | (103.43) |  | (0.03) |
| Marginalised | 1158.17*** | Marginalised | 2.87*** |
|  | (116.72) |  | (0.04) |
| Black | 849.00*** | Intercept | 2.41*** |
|  | (105.16) |  | (0.07) |
| Photo or Video | 533.19** | Black | 1.35*** |
|  | (175.53) |  | (0.03) |
| Date | 0.91*** | Date | 0.00*** |
|  | (0.17) |  | (0.00) |
| Text Length | -0.60*** | Text Length | 0.00*** |
|  | (0.09) |  | (0.00) |
| Sentiment | -153.79** | Sentiment | -0.08*** |
|  | (49.94) |  | (0.02) |
| Link | -752.33*** | Photo or Video | -0.19*** |
|  | (163.40) |  | (0.05) |
| Intercept | -1033.21*** | Link | -0.84*** |
|  | (245.00) |  | (0.05) |
| N | 67502 | N | 67502 |
| Adjusted $R^2$ | 0.01 | Adjusted $R^2$ | 0.18 |
| Log-Likelihood | -698660 | Log-Likelihood | -150830 |

*$p < 0.1$; **$p < 0.05$; ***$p < 0.01$

**Supplementary Table 2.** OLS linear regression models for Facebook, using the audiences variables. Dependent variables are engagement (left) and Box-Cox transformed engagement (right). Results shown in descending order of regression coefficient.

|  | Engagement |  | Engagement |
| --- | --- | --- | --- |
| Right | 3.52*** | Right | 3.53*** |
|  | (0.00) |  | (0.01) |
| Intercept | 3.28*** | Intercept | 3.42*** |
|  | (0.00) |  | (0.03) |
| Marginalised | 2.91*** | Marginalised | 2.80*** |
|  | (0.00) |  | (0.02) |
| Black | 2.63*** | Black | 2.57*** |
|  | (0.00) |  | (0.01) |
| Photo or Video | 0.43*** | Photo or Video | 0.41*** |
|  | (0.00) |  | (0.02) |
| Date | 0.00*** | Date | 0.00*** |
|  | (0.00) |  | (0.00) |
| Text Length | 0.00*** | Text Length | 0.00*** |
|  | (0.00) |  | (0.00) |
| Sentiment | -0.14*** | Sentiment | -0.18*** |
|  | (0.00) |  | (0.01) |
| Link | -1.47*** | Link | -1.52*** |
|  | (0.00) |  | (0.02) |
| N | 67502 | N | 67502 |
| (Pearson $X^2$) / (Df Residuals) | 43560.07 | (Pearson $X^2$) / (Df Residuals) | 48.30 |
| Log Likelihood | -108180000 | Log Likelihood | -520610 |

*$p < 0.1$; **$p < 0.05$; ***$p < 0.01$

**Supplementary Table 3.** GLM regression models for count data for Facebook, using the audiences variables. Dependent variables are engagement for the Poisson model (left) and the negative binomial model (right). Results shown in descending order of regression coefficient.

Supplementary Table 3 shows that the negative binomial regression improves on the goodness of fit (Pearson $X^2$) / (Df Residuals) metric by a factor of more than 900, compared to the Poisson regression (43560.07/48.3 = 901.86).

Instagram

|  | **Engagement** |  | **Ln(Engagement+0.5)** |
|---|---|---|---|
| Marginalised | 1880.67*** | Right | 2.28*** |
|  | (42.02) |  | (0.02) |
| Right | 1231.11*** | Marginalised | 2.14*** |
|  | (33.49) |  | (0.02) |
| Black | 676.78*** | Intercept | 1.55*** |
|  | (33.00) |  | (0.02) |
| Sentiment | 92.13*** | Black | 1.26*** |
|  | (16.72) |  | (0.01) |
| Number of hashtags | 5.58*** | Sentiment | 0.08*** |
|  | (1.11) |  | (0.01) |
| Date | 3.59*** | Number of hashtags | 0.03*** |
|  | (0.04) |  | (0.00) |
| Text length | -0.72*** | Date | 0.00*** |
|  | (0.03) |  | (0.00) |
| Intercept | -1531.08*** | Text length | 0.00*** |
|  | (39.19) |  | (0.00) |
| N | 116205 | N | 116205 |
| Adjusted $R^2$ | 0.11 | Adjusted $R^2$ | 0.46 |
| Log-Likelihood | -1108500 | Log-Likelihood | -209420 |

*$p < 0.1$; **$p < 0.05$; ***$p < 0.01$

**Supplementary Table 4.** OLS linear regression models for Instagram, using the audiences variables. Dependent variables are engagement (left) and Box-Cox transformed engagement (right). Results shown in descending order of regression coefficient.

Supplementary Table 4 shows that using the Box-Cox transformation of engagement as the outcome improves on the goodness of fit (Adjusted R2) by a factor of more than 4 (0.46 / 0.11 = 4.18).

|  | **Engagement** |  | **Engagement** |
| --- | ---: | --- | ---: |
| Intercept | 1.53*** | Intercept | 1.73*** |
|  | (0.00) |  | (0.01) |
| Marginalised | 1.22*** | Right | 1.70*** |
|  | (0.00) |  | (0.01) |
| Right | 0.88*** | Marginalised | 1.20*** |
|  | (0.00) |  | (0.01) |
| Black | 0.11*** | Black | 0.02*** |
|  | (0.00) |  | (0.01) |
| Sentiment | 0.01*** | Sentiment | 0.01*** |
|  | (0.00) |  | (0.01) |
| Number of hashtags | 0.00*** | Number of hashtags | 0.00*** |
|  | (0.00) |  | (0.00) |
| Date | 0.00*** | Date | 0.00*** |
|  | (0.00) |  | (0.00) |
| Text length | 0.00*** | Text length | 0.00*** |
|  | (0.00) |  | (0.00) |
| N | 116205 | N | 116205 |
| (Pearson $X^2$) / (Df Residuals) | 4544.00 | (Pearson $X^2$) / (Df Residuals) | 2.87 |
| Log Likelihood | -126740000 | Log Likelihood | -918920 |

*$p < 0.1$; **$p < 0.05$; ***$p < 0.01$

**Supplementary Table 5.** GLM regression models for count data for Instagram, using the audiences variables. Dependent variables are engagement for the Poisson model (left) and the negative binomial model (right). Results shown in descending order of regression coefficient.

Supplementary Table 5 shows that the negative binomial regression improves on the goodness of fit (Pearson $X^2$) / (Df Residuals) metric by a factor of more than 1500, compared to the Poisson regression (4544 .00/ 2.87 = 1583).

## Sub-audiences

Facebook

|  | Engagement |  | Engagement |
|---|---|---|---|
| Marginalised – Latin American | 2490.59*** <br> (638.21) | Marginalised – LGBT | -147.08 <br> (597.72) |
| Right – Nativist | 2046.55*** <br> (602.15) | Right – Veterans & Police | -498.31 <br> (613.88) |
| Right – Southern Identity | 1441.21** <br> (599.03) | Black – Identity | -816.75 <br> (598.26) |
| Right – Christian | 1125.84* <br> (624.28) | Right – Pro-Trump | -819.65 <br> (2727.10) |
| Photo or video | 653.79*** <br> (176.43) | Black – Religion | -845.41 <br> (677.90) |
| Right – Patriotic | 564.18 <br> (596.50) | Other – Liberal | -907.90 <br> (615.66) |
| Black – Social Justice & Activism | 541.82 <br> (595.76) | Marginalised – Native American | -989.96 <br> (643.26) |
| Right – Second Amendment | 332.12 <br> (633.78) | Right – Anti-Clinton | -1093.68 <br> (827.45) |
| Marginalised – Muslim | 95.14 <br> (603.10) | Other – Internet Culture | -1118.26* <br> (602.27) |
| Sentiment | 5.59 <br> (51.03) | Link | -1119.00*** <br> (166.66) |
| Date | 0.86*** <br> (0.18) | Black – Self-Defense | -1155.69* <br> (626.67) |
| Text length | -0.72*** <br> (0.09) | Black – Other | -1207.87 <br> (1519.75) |
| Intercept | -70.41 <br> (620.99) | Right – Internet Culture & Alt Right | -1230.19 <br> (794.06) |
| N | 67502 |  |  |
| Adjusted $R^2$ | 0.02 |  |  |
| Log-Likelihood | -698000 |  |  |

*$p < 0.1$; **$p < 0.05$; ***$p < 0.01$

**Supplementary Table 6.** OLS linear regression model for Facebook, using the sub-audiences variables. Dependent variable is engagement. Results shown in descending order of regression coefficient

|                              | Ln (Engagement +0.5) |                              | Ln (Engagement +0.5) |
|------------------------------|----------------------|------------------------------|----------------------|
| Marginalised – Latin American | 6.13***             | Other – Internet Culture     | 1.62***              |
|                              | (0.17)               |                              | (0.16)               |
| Right – Christian            | 5.96***              | Right – Pro-Trump            | 1.27*                |
|                              | (0.16)               |                              | (0.71)               |
| Right – Southern Identity    | 5.93***              | Black – Self-Defense         | 0.45**               |
|                              | (0.16)               |                              | (0.16)               |
| Right – Nativist             | 5.81***              | Intercept                    | 0.13                 |
|                              | (0.16)               |                              | (0.16)               |
| Right – Second Amendment     | 5.41***              | Sentiment                    | 0.10***              |
|                              | (0.17)               |                              | (0.01)               |
| Marginalised – LGBT          | 5.01***              | Photo or video               | 0.06                 |
|                              | (0.16)               |                              | (0.05)               |
| Marginalised – Muslim        | 4.85***              | Right – anti Clinton         | 0.05                 |
|                              | (0.16)               |                              | (0.22)               |
| Black – Social Justice & Activism | 4.50***         | Date                         | 0.00***              |
|                              | (0.16)               |                              | (0.00)               |
| Right – Patriotic            | 4.42***              | Text length                  | 0.00***              |
|                              | (0.16)               |                              | (0.00)               |
| Right – Veterans & Police    | 3.53***              | Right – Internet Culture & Alt Right | -0.90***     |
|                              | (0.16)               |                              | (0.21)               |
| Other – Liberal              | 2.93***              | Black – Religion             | -1.09***             |
|                              | (0.16)               |                              | (0.18)               |
| Black – Identity             | 2.62***              | Link                         | -1.27***             |
|                              | (0.16)               |                              | (0.04)               |
| Marginalised – Native American | 2.46***            | Black – Other                | -1.62***             |
|                              | (0.17)               |                              | (0.40)               |
| N                            | 67502                |                              |                      |
| Adjusted $R^2$               | 0.38                 |                              |                      |
| Log-Likelihood               | -142000              |                              |                      |

*$p < 0.1$; **$p < 0.05$; ***$p < 0.01$

**Supplementary Table 7.** OLS linear regression model for Facebook, using the sub-audiences variables. Dependent variable is Box-Cox transformed engagement. Results shown in descending order of regression coefficient.

Supplementary Tables 6 and 7 show that using the Box-Cox transformation of engagement as the outcome improves on the goodness of fit (Adjusted $R^2$) by a factor of 19 (0.38 / 0.02 = 19).

|  | Engagement |  | Engagement |
| --- | --- | --- | --- |
| Right – Nativist | 7.82*** | Marginalised – Native American | 3.71*** |
|  | (0.10) |  | (0.10) |
| Marginalised – Latin American | 7.81*** | Other – Internet Culture | 3.50*** |
|  | (0.10) |  | (0.10) |
| Right – Southern Identity | 7.52*** | Black – Self-Defense | 2.46*** |
|  | (0.10) |  | (0.10) |
| Right – Christian | 7.35*** | Right – anti Clinton | 0.94*** |
|  | (0.10) |  | (0.11) |
| Black – Social Justice & Activism | 7.07*** | Photo or video | 0.51*** |
|  | (0.10) |  | (0.00) |
| Right – Patriotic | 7.07*** | Right – Internet Culture & Alt Right | 0.00 |
|  | (0.10) |  | (0.12) |
| Right – Second Amendment | 6.90*** | Date | 0.00*** |
|  | (0.10) |  | (0.00) |
| Marginalised – Muslim | 6.67*** | Text length | 0.00*** |
|  | (0.10) |  | (0.00) |
| Marginalised – LGBT | 6.32*** | Sentiment | -0.01*** |
|  | (0.10) |  | (0.00) |
| Right – Veterans & Police | 5.98*** | Black – Religion | -0.53*** |
|  | (0.10) |  | (0.11) |
| Black – Identity | 4.90*** | Intercept | -0.54*** |
|  | (0.10) |  | (0.10) |
| Other – Liberal | 4.33*** | Link | -1.76*** |
|  | (0.10) |  | (0.00) |
| Right – Pro-Trump | 3.87*** | Black – Other | -2.03*** |
|  | (0.11) |  | (0.42) |
| N | 67502 |  |  |
| (Pearson $X^2$) / (Df Residuals) | 32752 |  |  |
| Log-Likelihood | -92300000 |  |  |

*$p < 0.1$; ** $p < 0.05$; *** $p < 0.01$*

**Supplementary Table 8.** Poisson regression model for count data for Facebook, using the sub-audiences variables. Results shown in descending order of regression coefficient

|  | Engagement |  | Engagement |
| --- | --- | --- | --- |
| Right - Nativist | 7.52* | Marginalised - Native American | 3.51* |
|  | (0.12) |  | (0.12) |
| Marginalised - Latin American | 7.52* | Other - Internet Culture | 3.26* |
|  | (0.12) |  | (0.12) |
| Right - Southern Identity | 7.23* | Black - Self-Defense | 2.68* |
|  | (0.12) |  | (0.12) |
| Right - Christian | 7.05* | Right - anti Clinton | 0.69* |
|  | (0.12) |  | (0.15) |
| Right - Patriotic | 6.90* | Right - Internet Culture Alt Right | -0.22 |
|  | (0.12) |  | (0.15) |
| Black - Social Justice Activism | 6.75* | Black - Religion | -0.76* |
|  | (0.12) |  | (0.14) |
| Right - Second Amendment | 6.65* | Black - Other | -2.26* |
|  | (0.12) |  | (0.46) |
| Marginalised - Muslim | 6.28* | Date | 0.0003* |
|  | (0.12) |  | (0.00002) |
| Marginalised - LGBT | 5.97* | Text length | -0.0008* |
|  | (0.12) |  | (0.00001) |
| Right - Veterans & Police | 5.71* | Sentiment | -0.0486* |
|  | (0.12) |  | (0.007) |
| Right - pro Trump | 5.35* | Photo/video | -0.16* |
|  | (0.37) |  | (0.02) |
| Black - Identity | 4.57* | Link | -1.39* |
|  | (0.12) |  | (0.02) |
| Other - Liberal | 4.05* | Intercept | 0.70* |
|  | (0.12) |  | (0.12) |
| N | 67502 |  |  |
| (Pearson $X^2$) / (Df Residuals) | 56.91 |  |  |
| Log-Likelihood | -496000 |  |  |

*$p < 0.1$; **$p < 0.05$; ***$p < 0.01$

**Supplementary Table 9.** Negative binomial regression model for count data for Facebook, using the sub-audiences variables. Results shown in descending order of regression coefficient

Supplementary Tables 8 and 9 show that the negative binomial regression improves on the goodness of fit (Pearson $X^2$) / (Df Residuals) metric by a factor of more than 560, compared to the Poisson regression (32752 / 56.91 = 575.51).

Instagram

|  | **Engagement** |  | **Engagement** |
| --- | --- | --- | --- |
| Marginalised – Women | 5570.97*** | Marginalised – Muslim | 297.41*** |
|  | (77.94) |  | (63.25) |
| Marginalised – LGBT | 3192.96*** | Other – Liberal | 275.73*** |
|  | (67.22) |  | (74.53) |
| Right – Veterans & Police | 2686.64*** | Right – Second Amendment | 138.88* |
|  | (48.67) |  | (81.32) |
| Right – Christian | 1842.39*** | Black – Social Justice & Activism | 131.77*** |
|  | (63.78) |  | (40.04) |
| Marginalised – Latin American | 1345.95*** | Date | 3.50*** |
|  | (74.47) |  | (0.04) |
| Right – Patriotic | 1313.05*** | Text length | -0.62*** |
|  | (41.08) |  | (0.03) |
| Black – Identity | 1200.60*** | Number of hashtags | -6.05*** |
|  | (36.56) |  | (1.12) |
| Other – Local Issues | 1013.43*** | Sentiment | -16.30 |
|  | (230.28) |  | (16.44) |
| Right – Internet Culture & Alt Right | 963.97*** | Marginalised – Native American | -151.38* |
|  | (68.14) |  | (84.63) |
| Black – Other | 720.52 | Black – Self-Defense | -260.44*** |
|  | (868.17) |  | (81.19) |
| Right – Nativist | 438.90*** | Intercept | -1430.55*** |
|  | (53.15) |  | (40.42) |
| Right – Southern Identity | 413.84*** |  |  |
|  | (58.40) |  |  |
| N | 116205 |  |  |
| Adjusted $R^2$ | 0.17 |  |  |
| Log-Likelihood | -1100000 |  |  |

*$p < 0.1$; **$p < 0.05$; ***$p < 0.01$*

**Supplementary Table 10.** OLS linear regression model for Instagram, using the sub-audiences variables. Dependent variable is engagement. Results shown in descending order of regression coefficient

|  | Ln(Engagement+0.5) |  | Ln(Engagement+0.5) |
| --- | --- | --- | --- |
| Marginalised – LGBT | 3.43*** | Right – Second Amendment | 1.56*** |
|  | (0.03) |  | (0.04) |
| Marginalised – Women | 3.04*** | Marginalised – Native American | 1.41*** |
|  | (0.03) |  | (0.04) |
| Right – Veterans & Police | 2.92*** | Black – Social Justice & Activism | 1.39*** |
|  | (0.02) |  | (0.02) |
| Right – Patriotic | 2.91*** | Intercept | 1.20*** |
|  | (0.02) |  | (0.02) |
| Right – Christian | 2.90*** | Black – Self-Defense | 0.82*** |
|  | (0.03) |  | (0.04) |
| Right – Internet Culture & Alt Right | 2.58*** | Black – Other | 0.62 |
|  | (0.03) |  | (0.37) |
| Right – Nativist | 2.34*** | Sentiment | 0.04*** |
|  | (0.02) |  | (0.01) |
| Other – Liberal | 2.31*** | Number of hashtags | 0.03*** |
|  | (0.03) |  | (0.00) |
| Right – Southern Identity | 2.20*** | Date | 0.00*** |
|  | (0.03) |  | (0.00) |
| Marginalised – Muslim | 2.19*** | Text length | 0.00* |
|  | (0.03) |  | (0.00) |
| Marginalised – Latin American | 2.17*** | Other – Local Issues | -0.74*** |
|  | (0.03) |  | (0.10) |
| Black – Identity | 1.85*** |  |  |
|  | (0.02) |  |  |
| N | 116205 |  |  |
| Adjusted $R^2$ | 0.51 |  |  |
| Log-Likelihood | -203000 |  |  |

*$p < 0.1$; **$p < 0.05$; ***$p < 0.01$

**Supplementary Table 11.** OLS linear regression model for Instagram, using the sub-audiences variables. Dependent variable is Box-Cox transformed engagement. Results shown in descending order of regression coefficient

Supplementary Tables 6 and 7 show that using the Box-Cox transformation of engagement as the outcome improves on the goodness of fit (Adjusted $R^2$) by a factor of 3 (0.51/ 0.17= 3).

|  | Engagement |  | Engagement |
| --- | --- | --- | --- |
| Intercept | 3.68*** | Right – Second Amendment | 0.81*** |
|  | (0.00) |  | (0.00) |
| Marginalised – Women | 2.65*** | Marginalised – Muslim | 0.63*** |
|  | (0.00) |  | (0.00) |
| Marginalised – LGBT | 2.34*** | Black – Social Justice & Activism | 0.41*** |
|  | (0.00) |  | (0.00) |
| Right – Veterans & Police | 1.97*** | Marginalised – Native American | 0.21*** |
|  | (0.00) |  | (0.00) |
| Right – Christian | 1.84*** | Sentiment | 0.02*** |
|  | (0.00) |  | (0.00) |
| Right – Patriotic | 1.53*** | Date | 0.00*** |
|  | (0.00) |  | (0.00) |
| Marginalised – Latin American | 1.51*** | Number of hashtags | 0.00*** |
|  | (0.00) |  | (0.00) |
| Black – Identity | 1.42*** | Text length | 0.00*** |
|  | (0.00) |  | (0.00) |
| Right – Internet Culture & Alt Right | 1.32*** | Black – Self-Defense | -0.57*** |
|  | (0.00) |  | (0.00) |
| Other – Liberal | 0.99*** | Black – Other | -1.06*** |
|  | (0.00) |  | (0.05) |
| Right – Southern Identity | 0.99*** | Other – Local Issues | -2.03*** |
|  | (0.00) |  | (0.02) |
| Right – Nativist | 0.84*** |  |  |
|  | (0.00) |  |  |
| N | 116205 |  |  |
| (Pearson $X^2$) / (Df Residuals) | 3555 |  |  |
| Log-Likelihood | -104000000 |  |  |

*$p < 0.1$; **$p < 0.05$; ***$p < 0.01$*

**Supplementary Table 12.** Poisson regression model for Instagram, using the sub-audiences variables. Dependent variable is engagement. Results shown in descending order of regression coefficient

|                                  | Engagement      |                              | Engagement      |
| -------------------------------- | --------------- | ---------------------------- | --------------- |
| Marginalised - LGBT              | 2.62*** (0.02)  | Black - Social Justice & Activism | 1.16*** (0.01)  |
| Marginalised - Women             | 2.54*** (0.02)  | Right - Second Amendment     | 0.91*** (0.03)  |
| Right - Veterans & Police        | 2.49*** (0.01)  | Marginalised - Native American | 0.43*** (0.03)  |
| Right - Patriotic                | 2.21*** (0.01)  | Black - Self-Defense         | -0.16*** (0.03) |
| Right - Religious                | 2.11*** (0.02)  | Black - Other                | -0.22 (0.28)    |
| Black - Identity                 | 1.71*** (0.01)  | Other - Local Interest       | -1.25*** (0.08) |
| Right - Internet Culture Alt Right | 1.54*** (0.02) | Number of hashtags           | 0.008*** (0)    |
| Marginalised - Latin American    | 1.53*** (0.02)  | Date                         | 0.004*** (0.00001) |
| Right - Southern Identity        | 1.32*** (0.02)  | Text Length                  | -0.00007*** (0.000009) |
| Marginalised - Muslim            | 1.31*** (0.02)  | Sentiment                    | -0.05*** (0.01) |
| Right - Nativist                 | 1.26*** (0.02)  | Intercept                    | 2.65*** (0.01)  |
| Other - Liberal                  | 1.19*** (0.02)  |                              |                 |
| N                                | 116205          |                              |                 |
| (Pearson $X^2$) / (Df Residuals) | 2.53            |                              |                 |
| Log-Likelihood                   | -906150         |                              |                 |

*$p < 0.1$; **$p < 0.05$; ***$p < 0.01$*

**Supplementary Table 13.** Negative binomial regression model for Instagram, using the sub-audiences variables. Dependent variable is engagement. Results shown in descending order of regression coefficient

Supplementary Tables 12 and 13 show that the negative binomial regression improves on the goodness of fit (Pearson $X^2$) / (Df Residuals) metric by a factor of more than 1400, compared to the Poisson regression (3555 / 2.53 = 1405.14 ).

## S.3. Nonlinear regression results for Facebook and Instagram

As the linear models impose parametric restrictions in the functional form of the relationship between the dependent variable and the independent variables (features), in order to ensure that out findings hold regardless of parametric assumption, we relax the linearity assumption and also implement nonlinear models. We use nonlinear decision-tree ensemble learning regression models, specifically Random Forests and Gradient Boosting.

Their results are reported below, starting with the feature combination that includes audience variables, and proceeding to the feature combination that instead uses sub-audience variables, in each case reporting Facebook results first followed by Instagram results.

**Audiences**

Facebook

|  | **Random Forests** | **Gradient Boosting** |
|---|---|---|
| Average $R^2$ | 0.01 | -0.002 |
| Maximum $R^2$ (best fold) | 0.05 | 0.06 |
| Minimum $R^2$ (worst fold) | 0.003 | -0.07 |
| **Baseline** | | |
| Maximum $R^2$ (best of ten folds) | -0.0006 | -0.0006 |
| $R^2$ for a single fold | -0.0001 | -0.0001 |

**Supplementary Table 14.** Facebook with audience variables: $R^2$ goodness of fit metrics, for Random Forests and Gradient Boosting regressors using 10-fold cross-validation. For each algorithm, the respective baseline goodness of fit metrics are reported in the lower half of the table, for two baselines: one using 10-fold cross validation, for which the maximum is reported (i.e. $R^2$ of best fold), and another baseline that does not split data into folds (one single fold).

The performance metrics of the baselines are very close to zero (when rounded to two decimal places), but actually slightly negative. We note that, for these nonlinear algorithms, it is possible for the $R^2$ score to be negative.[1] This means that the baseline decision tree algorithms, which use the input variables to construct the decisions trees which partition the value space of the input variables, but predict the mean at each partition, preform slightly worse than disregarding all input features and predicting the mean. And indeed, the engagement variable is very skewed, so this is a hard prediction task. So in this context the Maximum $R^2$ values of the algorithms represent a substantial improvement upon the baseline.

---

[1] As stated in the Python documentation at https://scikit-learn.org/stable/modules/model_evaluation.html#r2-score (Accessed 1 November 2019): "A constant model that always predicts the expected value of y, disregarding the input features, would get a R² score of 0.0."

|  | **Partial Dependence** |
| --- | --- |
| **Binary features** | |
| Right | 1143.71 |
| Marginalised | 254.88 |
| Black | 179.90 |
| Photo or video | 213.71 |
| Link | -854.23 |
| **Non-binary features** | |
| Date | 314.19† |
| Text length | -129.18† |
| Sentiment | -1239.71† |

† *Naively calculated partial dependence for non-binary variable, by taking the slope of the partial dependence curve between the maximum and minimum values of the variable.*

**Supplementary Table 15.** Facebook with audience variables: Results for the best fold of the Gradient Boosting method ($R^2 = 0.06$). Partial dependence values of engagement on each feature.

Instagram

|  | Random Forests | Gradient Boosting |
|---|---|---|
| Average $R^2$ | 0.18 | 0.30 |
| Maximum $R^2$ (best fold) | 0.22 | 0.33 |
| Minimum $R^2$ ( worst fold) | 0.13 | 0.23 |
| **Baseline** |  |  |
| Maximum $R^2$ (best of ten folds) | -0.0001 | -0.00003 |
| $R^2$ for a single fold | -0.0001 | -0.0001 |

**Supplementary Table 16.** Instagram with audience variables: $R^2$ goodness of fit metrics, for Random Forests and Gradient Boosting regressors using 10-fold cross-validation. For each algorithm, the respective baseline goodness of fit metrics are reported in the lower half of the table, for two baselines: one using 10-fold cross validation, for which the maximum is reported (i.e. $R^2$ of best fold), and another baseline that does not split data into folds (one single fold).

|  | Partial Dependence |
|---|---|
| **Binary features** |  |
| Marginalised | 1552.81 |
| Right | 789.83 |
| Black | 452.09 |
| **Non-binary features** |  |
| Sentiment | 155.91[†] |
| Number of hashtags | 17.12[†] |
| Date | 2.56[†] |
| Text length | -0.74[†] |

[†] *Naively calculated partial dependence for non-binary variable, by taking the slope of the partial dependence curve between the maximum and minimum values of the variable.*

**Supplementary Table 17.** Instagram with audience variables: Results for the best fold of the Gradient Boosting method ($R^2 = 0.33$). Partial dependence values of engagement on each feature.

**Sub-audiences**

Facebook

|  | **Random Forests** | **Gradient Boosting** |
|---|---|---|
| Average $R^2$ | 0.02 | 0.04 |
| Maximum $R^2$ (best fold) | 0.07 | 0.07 |
| Minimum $R^2$ (worst fold) | 0.004 | 0.01 |
| **Baseline** | | |
| Maximum $R^2$ (best of ten folds) | -0.0006 | -0.0001 |
| $R^2$ for a single fold | -0.0001 | -0.0001 |

**Supplementary Table 18.** Facebook with sub-audience variables: $R^2$ goodness of fit metrics, for Random Forests and Gradient Boosting regressors using 10-fold cross-validation. For each algorithm, the respective baseline goodness of fit metrics are reported in the lower half of the table, for two baselines: one using 10-fold cross validation, for which the maximum is reported (i.e. $R^2$ of best fold), and another baseline that does not split data into folds (one single fold).

|  | Partial Dependence |  |
| --- | --- | --- |
| **Binary variables** |  |  |
| Marginalised – Latin American | 2071.97 |  |
| Right – Nativist | 1772.35 |  |
| Right – Southern Identity | 1470.67 |  |
| Right – Christian | 903.07 |  |
| Black – Social Justice & Activism | 588.99 |  |
| Right – Patriotic | 551.69 |  |
| Photo or video | 271.64 |  |
| Right – Second Amendment | 161.14 |  |
| Black – Other | 0.00 |  |
| Marginalised – LGBT | 0.00 |  |
| Marginalised – Muslim | 0.00 |  |
| Right – Pro-Trump | 0.00 |  |
| Right – Veterans & Police | -160.37 |  |
| Right – anti Clinton | -187.98 |  |
| Black – Religion | -241.50 |  |
| Right – Internet Culture & Alt Right | -363.11 |  |
| Marginalised – Native American | -481.73 |  |
| Other – Internet Culture | -500.47 |  |
| Black – Identity | -559.15 |  |
| Other – Liberal | -607.28 |  |
| Black – Self-Defense | -655.21 |  |
| Link | -835.06 |  |
| **Non-binary variables** |  | Adjusted $R^2$ |
| Date | 0.99 ‡ *** | 0.87 |
| Text length | -0.02 ‡ *** | 0.15 |
| Sentiment | -24.89 ‡ *** | 0.06 |

‡ *Partial dependence for non-binary variable, regression coefficient of this variable in the univariate OLS regression fitting on the partial dependence curve values of engagement on this variable. The adjusted $R^2$ for this regression is shown in the third column.*

*$p < 0.1$; ** $p < 0.05$; *** $p < 0.01$

**Supplementary Table 19.** Facebook with sub-audience variables: Results for the best fold of the Gradient Boosting method ($R^2 = 0.07$). Partial dependence values of engagement on each feature.

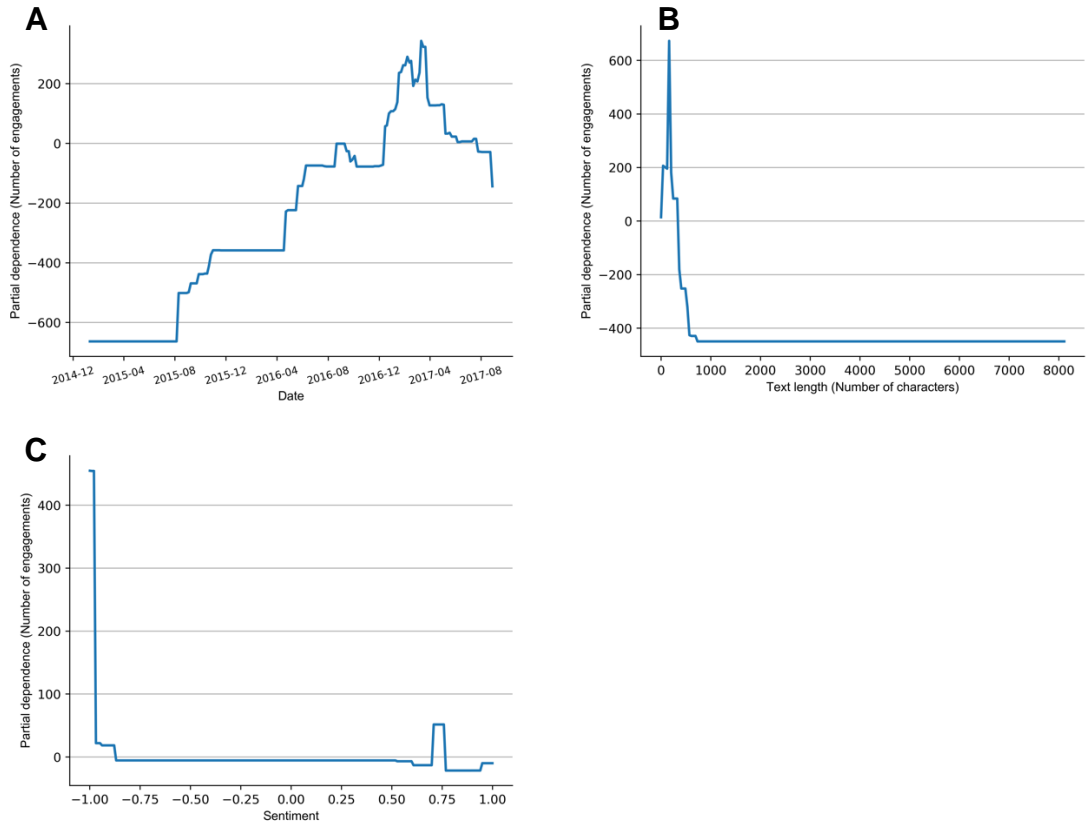

**Supplementary Figure 2**. Facebook with sub-audience variables: Partial dependence plots of engagement on Date (A), Text Length (B), and Sentiment (C), for the best fold of the Gradient Boosting regression ($R^2 = 0.07$).

Instagram

|  | **Random Forests** | **Gradient Boosting** |
|---|---|---|
| Average $R^2$ | 0.20 | 0.35 |
| Maximum $R^2$ (best fold) | 0.26 | 0.39 |
| Minimum $R^2$ (worst fold) | 0.14 | 0.29 |
| **Baseline** | | |
| Maximum $R^2$ (best of ten folds) | -0.0001 | -0.000004 |
| $R^2$ for a single fold | -0.0001 | -0.0001 |

**Supplementary Table 20.** Instagram with sub-audience variables: $R^2$ goodness of fit metrics, for Random Forests and Gradient Boosting regressors using 10-fold cross-validation. For each algorithm, the respective baseline goodness of fit metrics are reported in the lower half of the table, for two baselines: one using 10-fold cross validation, for which the maximum is reported (i.e. $R^2$ of best fold), and another baseline that does not split data into folds (one single fold).

|  | **Partial Dependence** |  |
| --- | --- | --- |
| **Binary variables** |  |  |
| Marginalised – Women | 4476.31 |  |
| Marginalised – LGBT | 2844.73 |  |
| Right – Veterans & Police | 2077.99 |  |
| Right – Christian | 1049.60 |  |
| Black – Identity | 810.45 |  |
| Right – Patriotic | 615.53 |  |
| Marginalised – Latin American | 614.83 |  |
| Right – Internet Culture & Alt Right | 144.79 |  |
| Black – Other | 0.00 |  |
| Marginalised – Muslim | 0.00 |  |
| Other – Liberal | 0.00 |  |
| Other – Local Issues | 0.00 |  |
| Right – Second Amendment | 0.00 |  |
| Right – Nativist | 0.00 |  |
| Right – Southern Identity | 0.00 |  |
| Marginalised – Native American | -178.00 |  |
| Black – Social Justice & Activism | -332.39 |  |
| Black – Self-Defense | -377.76 |  |
| **Non-binary variables** |  | **Adjusted $R^2$** |
| Number of hashtags | 0.2 ‡ | 0.02 |
| Sentiment | 18.80 ‡ *** | 0.58 |
| Date | 2.81 ‡ *** | 0.85 |
| Text length | -0.10 ‡ *** | 0.26 |

‡ *Partial dependence for non-binary variable, regression coefficient of this variable in the univariate OLS regression fitting on the partial dependence curve values of engagement on this variable. The adjusted $R^2$ for this regression is shown in the third column.*

*\* $p < 0.1$; \*\* $p < 0.05$; \*\*\* $p < 0.01$*

**Supplementary Table 21.** Instagram with sub-audience variables: Results for the best fold of the Gradient Boosting method ($R^2 = 0.39$). Partial dependence values of engagement on each feature.

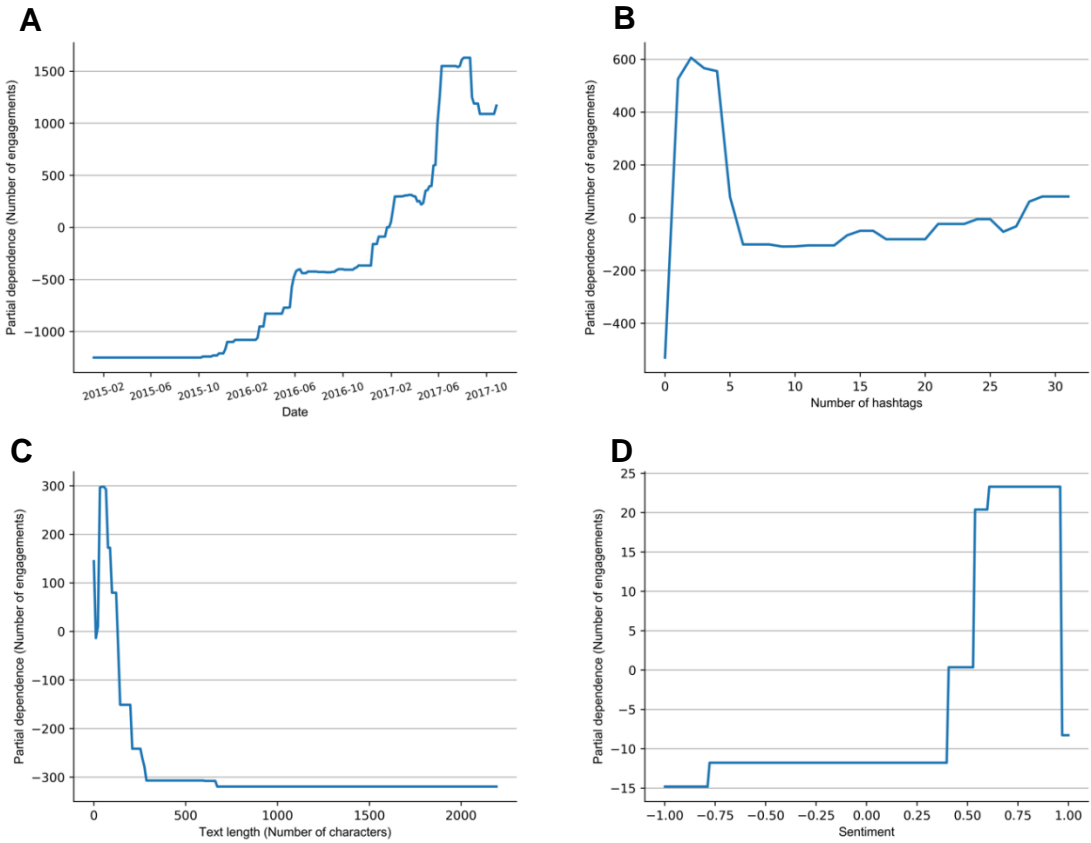

**Supplementary Figure 3.** Instagram with sub-audience variables: Partial dependence plots of engagement on the date (A); number of hashtags (B); text length (C), and sentiment (D), for the best fold of the Gradient Boosting regression ($R^2 = 0.07$).